\documentclass[fleqn,usenatbib]{mnras}

\usepackage{newtxtext,newtxmath}
\usepackage[T1]{fontenc}

\DeclareRobustCommand{\VAN}[3]{#2}
\let\VANthebibliography\thebibliography
\def\thebibliography{\DeclareRobustCommand{\VAN}[3]{##3}\VANthebibliography}

\usepackage{graphicx}	% Including figure files
\usepackage{amsmath}% Advanced maths commands
\usepackage{color}
\usepackage{tabularx}
\title{The \ion{H}{i} Gas Disk Thickness of the Ultra-diffuse Galaxy AGC 242019}
\author[]{
Xin Li,$^{1}$
Yong Shi,$^{1,2}$\thanks{E-mail: yong@nju.edu.cn}
Zhi-Yu Zhang,$^{1,2}$
Jianhang Chen,$^{3}$
Xiaoling Yu,$^{1}$
Junzhi Wang,$^{4}$
Qiusheng Gu$^{1,2}$
\newauthor
and Songlin Li$^{5,6}$
\\
$^{1}$School of Astronomy and Space Science, Nanjing University, Nanjing 210093, People’s Republic of China\\
$^{2}$Key Laboratory of Modern Astronomy and Astrophysics (Nanjing University), Ministry of Education, Nanjing 210093, People’s Republic of China\\
$^{3}$European Southern Observatory, Karl-Schwarzschild-Strasse 2, D-85748 Garching bei Muenchen, Germany\\
$^{4}$Shanghai Astronomical Observatory, Chinese Academy of Sciences, 80 Nandan Road, Shanghai 200030, People’s Republic of China\\
$^{5}$Research School of Astronomy and Astrophysics, Australian National University, Canberra, ACT 2611, Australia\\
$^{6}$ARC Centre of Excellence for All Sky Astrophysics in 3 Dimensions (ASTRO 3D), Australia
}

\date{Accepted XXX. Received YYY; in original form ZZZ}
\pubyear{2021}

\begin{document}
\label{firstpage}
\pagerange{\pageref{firstpage}--\pageref{lastpage}}
\maketitle  
% \abstract{}{}{}{}{} 
% 5 {} token are mandatory
 
\begin{abstract}

Ultra-diffuse galaxies (UDGs) are as faint as dwarf galaxies but whose sizes are similar to those of spiral galaxies.  A variety of formation mechanisms have been proposed, some of which could result in different disk thicknesses. In this study, we measure the radial profile of the \ion{H}{i} scale height ($h_{\rm g}$) and flaring angle $( h_{\rm g}/R )$ of AGC 242019 through the joint Poisson-Boltzmann equation based on its well spatially-resolved HI gas maps. The mean \ion{H}{i} scale height of AGC 242019 is $\left \langle h_{\rm g} \right \rangle \approx$ $537.15 \pm 89.4$ pc, and the mean flaring angle is $\left \langle h_{\rm g}/R \right \rangle \approx$ $0.19 \pm 0.03$. As a comparison, we also derive the disk thickness for a sample of 14 dwarf irregulars. It is found that the \ion{H}{i} disk of AGC 242019 has comparable thickness to dwarfs. This suggests that AGC 242019 is unlikely to experience much stronger stellar feedback than dwarf galaxies, which otherwise leads to a thicker disk for this galaxy.

\end{abstract}


\section{Introduction}

%latexdiff -p preamble.tex main_S\ion{H}{i}.tex main.tex \right \rangle ms_diff.tex
%pdflatex ms_diff.tex
   
Ultra-diffuse  galaxies (UDGs)  are  characterized  as faint but extended  objects.  Their  central surface brightness is fainter than 24 mag/arcsec$^{2}$ while the  effective  radius  in the optical is larger than  1.5  kpc \citep{van_Dokkum2015}.   They have  been  found in  a  wide range  of environments. UDGs in  galaxy  clusters  are  red and  round  with negligible star formation \citep{van_Dokkum2015,Van_Dokkum2016}.  Some UDGs in  the clusters are suggested  to be failed  $L_{\rm \ast}$ galaxies, such as  Dragonfly 44  because of its  abundant global  clusters, high internal velocity  dispersion, and  massive host  halo.   Some  other cluster UDGs are  instead lacking of dark matter such as NGC1052-DF2 and NGC1052-DF4 \citep{van_Dokkum2018, van_Dokkum2019}.  These objects may lose  a  large   fraction  of  baryons  and  dark   matter  by  tidal interactions/stripping. In contrast to the cluster UDGs, UDGs in the field are gas rich  with  mild star formation \citep[][]{Leisman2017}. They likely reside in dwarf-size halos as indicated by their kinematic data \citep{Shi2021, ManceraPina2022}.

To further understand what mechanism regulates the UDG formation, it is important to characterize the disk thickness of UDGs.  As suggested by \cite{Chan2018}, the field UDGs with high spins tend to be more disky, while those in normal spin halos are puffier. \cite{Di_Cintio2017} pointed out that episodic stellar wind generated by bursty star formation can enlarge and  puff the disk of dwarf galaxies to produce UDGs. Cluster UDGs are more prolate and spheroidal \citep[e.g.][]{Burkert2017, Chan2018, Jiang2019, Liao2019}, and the thickness of their progenitors depends on their evolutionary channels. For example, progenitors of cluster UDGs with disks may become spheroidal throughout the tidal stirring \citep{Chan2018, Rong2020} or stellar feedback \citep[e.g.][]{Janssen2010, Liao2019, Bacchini2020(SN)}, while thicker and more prolate progenitors are formed by quenching their star formation through the ram-pressure stripping \citep{Chan2018}. 

For the gas-rich UDG, it is possible to estimate the thickness of gaseous disks through gas kinematic observations. As gas is collisional, shapes of gaseous disks are sensitive to galactic activities such as the gravity instability, stellar feedback, galaxy interaction, etc. \citep[e.g.][]{Scannapieco2008, Marinacci2017}. Here we quantify the thickness as the scale height, defined as the distance to the mid-plane whose volume density is half of the maximum along the vertical direction of the disk, and derive it theoretically based on vertical hydrostatic equilibrium. In a self-gravitating system, the disk stability is determined by gravity and pressure of vertical velocity dispersion \citep{Narayan2002a,Narayan2002b}. By adopting Poisson's equation that describes the gravitational potential of galaxies, several works estimated the scale height of spiral and dwarf galaxies with different approximations and assumptions \citep[e.g.][]{Narayan2002a,Banerjee2011,Banerjee2013,Patra2019(a),Sarkar2019,Patra2020,Das2020}. However, these methods require high spatial-resolution observational data.

We focus on the \ion{H}{i} disk of AGC 242019 whose major axis is resolved into 14 beams with a beam size of 9."85 $\times$ 9."33 \citep{Shi2021}, which makes it possible to obtain the reliable estimate of its \ion{H}{i} disk height. The object has $M_{\rm \ion{H}{i}}=(8.51 \pm 0.36)\times10^8 M_{\rm \odot}$ and $M_{\rm \ast}=(1.37 \pm 0.05) \times 10^8 M_{\rm \odot}$. In \S~2, we describe our methods and input parameters that we used to derive the scale height of the \ion{H}{i} disk. In \S~3, we show the results of the \ion{H}{i} scale height of AGC 242019. We compare the \ion{H}{i} scale height and flaring angle of our target with those of dwarf irregulars, and discuss these results in \S~4.

\section{Methods}
\label{sec:2}
Atomic gas and stars are modeled as two axis-symmetric disks, while molecular gas is estimated to be negligible in this galaxy \citep[][]{Shi2021}. These two components are further assumed to be co-planar and co-center with their host dark matter halo. The disks are assumed to be isothermal with each component satisfying hydrostatic equilibrium in the vertical direction.

\subsection{Joint Poisson-Boltzmann equation}
We adopt a joint Poisson-Boltzmann equation \citep[][]{Banerjee2013,Patra2019(a),Sarkar2019, Patra2020} for the star and gas components at a radius $R$ in a cylindrical coordinate as:
\begin{equation}
    \begin{split}
    \left \langle\sigma _{\rm z}\right \rangle^2_{\rm i} \frac{\partial}{\partial z}(\frac{1}{\rho_{\rm i}}\frac{\partial \rho_{\rm i}}{\partial z }) = -4\pi {\rm G} (\rho_{\rm g} + \rho_{\rm \ast} +\rho _{\rm DM})+ \frac{1}{R} \frac{\partial}{\partial R}(R \frac{\partial \Phi _{\rm tot}}{\partial R}),
   \end{split}
\label{eq:final}
\end{equation}
\begin{equation}
    (R \frac{\partial \Phi _{\rm tot}}{\partial R})_{\rm R,z} =(v_{\rm rot}^2)_{\rm R,z}.
\label{eq:radial}
\end{equation}
This equation is the combination of the vertical hydrostatic equilibrium equation given as $ \frac{\partial }{\partial z}(\rho _{\rm i} \left \langle\sigma _{\rm z}^2\right \rangle_{\rm i})+\rho _{\rm i} \frac{\partial \Phi _{\rm tot}}{\partial z}=0 $, and Poisson's equation that describes the gravitational potential of a galaxy as $\frac{\partial ^2 \Phi _{\rm tot}}{\partial z^2} + \frac{1}{R} \frac{\partial}{\partial R}(R \frac{\partial \Phi _{\rm tot}}{\partial R}) = 4\pi {\rm G}(\rho_{\rm g}+\rho _{\rm \ast} +\rho _{\rm DM})$, where $i$ denotes either stars or \ion{H}{i} components, $\left \langle\sigma _{\rm z}\right \rangle_{\rm i}$ is the vertical velocity dispersion and is independent of $z$ under the assumption of isothermal disks \citep{Kruit1981(99)}, ${\rm G}$ is gravitational constant, $\rho_{\rm g}$ is the gas volume density, $\rho _{\rm \ast}$ is the stellar volume density, $\rho_{\rm DM}$ is the dark matter volume density which is a function of both $R$ and $z$. The total gravitational potential $\Phi _{\rm tot}$ includes stars, gas and dark matter, and its radial gradient is measured through the rotation curve independent of $z$ \citep[see][for more details]{Banerjee2011}. Some studies do not consider $\partial \Phi _{\rm tot}/\partial R$ for spiral galaxies at the flat part of the rotation curve where the radial gradient becomes negligible \citep[e.g.][]{Narayan2002a,Das2020}.

\subsection{Input parameters}

The input parameters to solve the above equations are: (1) the radial profile of the \ion{H}{i} surface density $\Sigma_{\rm g}$; (2) the radial profile of the \ion{H}{i} velocity dispersion $\left \langle\sigma_{\rm z}\right \rangle_{\rm g}$; (3) the circular velocity $v_{\rm rot}$ as a function of radius; (4) the radial profile of stellar mass surface density $\Sigma_{\rm \ast}$; (5) the radial profile of the stellar velocity dispersion $\left \langle\sigma_{\rm z}\right \rangle_{\rm \ast}$; (6) the radial profile of the dark matter density $\rho_{\rm DM}$.

For AGC 242019, measurements of these parameters are detailed in \cite{Shi2021} except for $\left \langle\sigma_{\rm z}\right \rangle_{\rm \ast}$ that is present at the end of the section. The first three were based on the \ion{H}{i} data that was acquired through the Very Large Array (VLA) with a velocity resolution of $\sim 7$ km s$^{-1}$ and a beam size of $9."85 \times 9."33$.  The stellar surface density profile $\Sigma_{\rm \ast}$ was obtained from $3.6 \rm{ \mu m}$ broad band image of Wide-field Infrared Survey Explorer (WISE) \citep{Wright2010} with Kroupa initial mass function (IMF) and the mass-to-light ratio of $\gamma _{\rm 3.6\mu m}=0.6$. We fit $\Sigma_{\ast}$ with a S$\rm \acute{e}$rsic profile, 
$\Sigma_{\ast}(R) = \Sigma_{\rm \ast,0}\exp{\left [ -(R/R_{\rm d})^{1/n}\right ]}$ \citep[][equation~(14)]{Graham2005}, to get the disk scale length $R_{\rm d} = 4.95$ kpc. The radial profile of the dark matter density is well fitted by a Navarro-Frenk-White model \citep[NFW,][]{Navarro1997}:
\begin{equation}
   \rho_{\rm DM}(R,z) = \frac{\rho_{\rm c}\delta_{\rm char}}{(\frac{\sqrt{R^2+z^2}}{R_{\rm s}})(1+\frac{\sqrt{R^2+z^2}}{R_{\rm s}})^2},
\label{Cusp_DM}
\end{equation}
where $\rho_{\rm c}$ is the present critical density, $\delta_{\rm char} = \frac{200c^3g}{3},g=\frac{1}{{\rm ln}(1+c)-c/(1+c)}, c =R_{\rm 200}/R_{\rm s}$. As derived in \citet{Shi2021}, the scale radius $R_{\rm s}$ is 33.3$\pm$9.1 kpc and $R_{\rm 200}$ is 65.0$\pm$7.4 kpc. The radial profiles of the remaining five parameters are listed in Table~\ref{tab:para1}.

\begin{table*}
	\centering
	\caption{Radial profiles of different physical parameters of AGC 242019. } 
	\label{tab:para1}
	\resizebox{0.6\textwidth}{!}{
	\begin{tabular}{lcccccc} % four columns, alignment for each
		\hline
		\hline
		$R$ & $\Sigma_{\rm g}$ & $\left \langle\sigma_{\rm z}\right \rangle_{\rm g}$ & $v_{\rm rot}$ & $\Sigma_{\rm \ast}$ & $\left \langle\sigma_{\rm z}\right \rangle_{\rm \ast}$ & $h_{\rm g}$\\
		(kpc) & (M$_{\rm \odot}$ pc$^{\rm -2}$) & (km s$^{\rm -1}$) & (km s$^{\rm -1}$) & (M$_{\rm \odot}$ pc$^{\rm -2}$) & (km s$^{\rm -1}$) &(pc)\\
		 (1) & (2) & (3) & (4) & (5) & (6) & (7)\\
		 \hline
        0.67 & 2.78 $\pm$ 0.130 & 4.5 $\pm$ 1.3 & 10.4 $\pm$ 1.2 & 1.90 $\pm$ 0.049 & 4.84 & 183.68 $\pm$ 69.97 \\
        2.02 & 3.07 $\pm$ 0.092 & 7.7 $\pm$ 1.3 & 14.5 $\pm$ 1.3 & 1.79 $\pm$ 0.030 & 4.22 & 465.80 $\pm$ 116.28 \\
        3.36 & 2.94 $\pm$ 0.063 & 7.7 $\pm$ 1.0 & 22.3 $\pm$ 1.4 & 1.40 $\pm$ 0.028 & 3.69 & 549.60 $\pm$ 109.77 \\
        4.71 & 2.83 $\pm$ 0.052 & 7.0 $\pm$ 1.6 & 29.4 $\pm$ 1.8 & 0.81 $\pm$ 0.026 & 3.22 & 633.88 $\pm$ 233.30 \\
        6.06 & 2.73 $\pm$ 0.045 & 6.5 $\pm$ 1.6 & 33.8 $\pm$ 1.6 & 0.30 $\pm$ 0.041 & 2.81 & 852.79 $\pm$ 338.98 \\		
        \hline 
	\end{tabular}}
	\begin{tabularx}{\textwidth}{X}
	(1) the radius to galactic center in kpc; (2) the gas mass surface density in ${\rm M_{\odot}\, pc^{-2}}$; (3) the \ion{H}{i} vertical velocity dispersion in ${\rm km\, s^{-1}}$; (4) the circular velocity in ${\rm km\, s^{-1}}$; (5) the stellar mass surface density in ${\rm M_{\odot}\,pc^{-2}}$; (6) the stellar vertical velocity dispersion in ${\rm km \, s^{-1}}$ (see main text for details); (7) the \ion{H}{i} scale height in pc (see main text for details). All data in columns (1)-(5) are obtained from \protect \cite{Shi2021}.
	\end{tabularx}
\end{table*}

For dwarf irregulars as a comparison to AGC 242019, we carry out similar calculations for a sample of 14 objects as listed in Table~\ref{tab:dwarf_para1}. Their radial profiles of the HI surface density, velocity dispersion, and circular velocity were derived by \cite{Iorio2017} based on the \ion{H}{i} data of the LITTLE THINGS project \citep{Hunter2012}.
Their stellar mass surface density radial profiles were given by \cite{Zhang2012}, for which we fit the same S$\rm \acute{e}$rsic profile to get their scale lengths ($R_{\rm d}$). To obtain the rotation curve of dark matter, we first run the \texttt{ROTMOD} task from \texttt{GIPSY} \citep{gipsy} to estimate the rotation curves of stellar and gas components, and
then subtract them from the observed rotation curve. As \cite{Oh2015} showed that the dark matter profiles of these irregulars can be described by pseudo-isothermal (ISO) models \citep{ISO}, we then fit the rotation curve to obtain the radial profiles of dark matter density through ISO models:
\begin{equation}
    \rho_{\rm DM}(R,z) = \frac{\rho_0}{1+\frac{R^2+z^2}{{R_{\rm c}}^2}},
\label{ISO_DM}
\end{equation}
where $\rho_0$ is the central density, and $R_{\rm c}$ is the core radius of the halo. We note that our results change little if adopting the NFW profile \citep{Patra2020}. Some dwarf irregulars can have bar structures or active star formation \citep[e.g.][]{Hunter2006,Hunter2011,Zhang2012,Johnson2015,Iorio2017,Patra2019(b)}. Since bars or violent star formation may induce non-circular motion to break the hydrostatic equilibrium assumption \citep{Johnson2015,Iorio2017,Patra2019(b)}, we exclude radial regions of optical bars for WLM, DDO 154, NGC 2366, DDO 126, and DDO 133 \citep[e.g.][]{Hunter2006,Hunter2011,Zhang2012}, as well as the radial region of the \ion{H}{i} bar for DDO 164 \citep{Johnson2015,Iorio2017,Patra2019(b)}. The derived disk scale lengths, parameters of dark matter models  and adopted bar radii are listed in Table~\ref{tab:dwarf_para1}.

\begin{table*}
	\caption{Derived properties of dwarf irregulars. }
	\label{tab:dwarf_para1}
	\resizebox{0.6\textwidth}{!}{
    \begin{tabular}{lccccccc}
         \hline
         \hline
         Galaxy & $R_{\rm d}$ & $R_{\rm bar}$ & $\rho_{\rm 0}$ & $R_{\rm c}$& $\left \langle\sigma_{\rm z}\right \rangle_{\rm \ast,0}$ & $\left \langle h_{\rm g}\right \rangle_R$ & $\left \langle h_{\rm g}\right \rangle_{R/R_{\rm d}}$\\
          & (kpc) & (kpc) & (M$_{\rm \odot}$ pc$^{\rm -3}$) & (kpc) & (km s$^{\rm -1}$)& (pc) & (pc)\\
          (1) & (2) & (3) & (4) & (5) & (6)& (7) & (8)\\
          \hline
            DDO 101 & 0.61 &  & 0.452 $\pm$ 0.075 & 0.455 $\pm$ 0.052 & 6.53 & 80.93 $\pm$ 13.10 & 57.55 $\pm$ 14.25 \\
            WLM & 1.24 & ${1.00}^{\rm a}$ & 0.042 $\pm$ 0.006 & 1.046 $\pm$ 0.139 &2.15 &  474.64 $\pm$ 39.16 & 317.89 $\pm$ 44.99 \\
            DDO 50 & 1.49 &  & 0.198 $\pm$ 0.091 & 0.351 $\pm$ 0.099 & 9.06 & 474.12 $\pm$ 45.85 & 215.87 $\pm$ 31.39 \\
            DDO 87 & 1.85 &  & 0.037 $\pm$ 0.012 & 1.283 $\pm$ 0.380 & 3.21 & 437.46 $\pm$ 50.45 & 244.22 $\pm$ 40.06 \\
            CVnIdwA & 1.23 &  & 0.016 $\pm$ 0.020 & 0.955 $\pm$ 0.368 & 1.78 & 252.50 $\pm$ 45.88 & 263.61 $\pm$ 35.26 \\
            DDO 52 & 1.11 &  & 0.030 $\pm$ 0.010 & 1.494 $\pm$ 0.366 & 4.61 & 455.63 $\pm$ 52.30 & 176.78 $\pm$ 54.30 \\
            UGC 8508 & 0.54 &  & 0.095 $\pm$ 0.040 & 0.705 $\pm$ 0.332 & 2.81 & 297.04 $\pm$ 46.48 & 233.57 $\pm$ 37.47 \\
            DDO 168 & 1.72 & ${1.00}^{\rm b}$ & 0.030 $\pm$ 0.006 & 1.943 $\pm$ 0.420 &4.71 &  464.04 $\pm$ 45.24 & 467.66 $\pm$ 82.87 \\
            DDO 154 & 0.93 & ${1.13}^{\rm a}$ & 0.041 $\pm$ 0.005 & 1.134 $\pm$ 0.093 &2.12 &  715.22 $\pm$ 41.08 &  \\
            NGC 2366 & 2.93 & ${4.00}^{\rm a}$ & 0.022 $\pm$ 0.003 & 2.033 $\pm$ 0.264 &2.34 &  751.22 $\pm$ 91.40 &  \\
            DDO 126 & 1.47 & ${0.97}^{\rm a}$ & 0.019 $\pm$ 0.005 & 1.736 $\pm$ 0.504 &2.20 &  564.00 $\pm$ 52.25 & 642.19 $\pm$ 66.51 \\
            DDO 133 & 1.14 & ${1.48}^{\rm a}$ & 0.088 $\pm$ 0.022 & 0.813 $\pm$ 0.168 &3.89 &  355.93 $\pm$ 51.78 &  \\
            DDO 216 & 0.93 &  & 0.041 $\pm$ 0.023 & 0.427 $\pm$ 0.176 & 2.56 & 335.42 $\pm$ 51.27 & 230.26 $\pm$ 25.04 \\
            DDO 53 & 1.70 &  & 0.049 $\pm$ 0.023 & 0.528 $\pm$ 0.155 & 1.86 & 386.41 $\pm$ 65.62 & 215.00 $\pm$ 26.30 \\   
         \hline
    \end{tabular}}
    \begin{tabularx}{\textwidth}{X}
             Columns: (1) names of dwarf irregulars; (2) the disk scale length in kpc; (3) the bar length in kpc; (4) the central density of dark matter in M$_{\rm \odot}$ pc$^{\rm -3}$; (5) the core radius of dark matter halo in kpc; (6) the central stellar vertical velocity dispersion in km s$^{\rm -1}$ ; (7) the mean \ion{H}{i} scale height ($0.67 < R < 6.06$ kpc) in pc; (8) the mean \ion{H}{i} scale height ($0.1R_{\rm d} < R < 1.2 R_{\rm d}$) in pc.\\
    \emph{Notes.}$^{\rm a}$ Data obtained from \cite{Hunter2006}; $^{\rm b}$ Data obtained form \cite{Patra2019(b)}.\\
    \end{tabularx}		
\end{table*}

For both AGC 242019 and the comparison dwarfs, it is extremely difficult to measure the stellar velocity dispersion as a function of radius. As a result we adopt the radial profile of the stellar vertical dispersion proposed by e.g. \cite{KS1}: $\left \langle\sigma_{\rm z}\right \rangle_{\rm \ast}(r) = \left \langle\sigma_{\rm z}\right \rangle_{\rm \ast,0}{\rm exp}(-r/2R_{\rm d})$, where $\left \langle\sigma_{\rm z}\right \rangle_{\rm \ast,0}$ is the central stellar velocity dispersion. The simplified hydrostatic equilibrium with the isothermal stellar disk gives $\left \langle\sigma_{\rm z}\right \rangle_{\rm \ast,0} = \sqrt{2\pi {\rm G} \Sigma_{\ast} z_{\rm e}}$, where $z_{\rm e}$ is defined for the stellar vertical density profile $\rho_{\rm z}(z) = {\rm sech^2} (\frac{z}{2z_{\rm e}})$ \citep[see][for more details]{Kruit1988}. By fitting the vertical brightness distributions of 33 edge-on spiral galaxies in \cite{Kregel2002} with sech$^2(\frac{z}{2z_{\rm e}})$ profiles, we derive $z_{\rm e} \sim R_{\rm d}/9.5$.  We note that although such estimates of the stellar velocity dispersion can have large uncertainties,  we will show later that the adopted $\left \langle\sigma_{\rm z}\right \rangle_{\rm \ast}$ have no significant effects on the \ion{H}{i} scale heights.

\subsection{Solving the equation}

To help numerically solve the Poisson-Boltzmann equation, we further fit the rotation curves of dwarf irregulars with the Brandt profiles \citep{Brandt1960} to obtain the gradient. For those whose rotation curves do not reach the flat stages linear functions were adopted \citep[see][for more details]{Patra2019(a),Patra2020}. Such fitting can also smooth the small fluctuation of the rotation curve to avoid the rapid change in the radial gradient.

Equation~(\ref{eq:final}) is a bivariate second-order partial differential equation with two variables $\rho_{\rm \ast}$ and $\rho_{\rm g}$. We adopt the same solving process as done in \citet{Banerjee2011}, \citet{Patra2019(a)}, \citet{Sarkar2019} and \citet{Patra2020}. The python package \texttt{scipy} can solve equation~(\ref{eq:final}) numerically with the 8-order Runge-Kutta method, and we can get both \ion{H}{i} and stellar vertical profile at a given radius simultaneously. We set the initial conditions as $(\rho_{\rm i})_{\rm z=0}=\rho_{\rm i,0}$ and $\frac{d\rho_{\rm i}}{dz}|_{z=0}=0$, where $\rho_{\rm i,0}$ is volume density at mid-plane. At first, we need to put a trial $\rho_{\rm i,0}$, then iteratively solve the equation until the solved $\rho_{\rm i}$ meets the criteria of $2\int_{0}^{\infty}\rho_{\rm i}dz=\Sigma_{\rm i}$, with $1\%$ differences \citep[see Appendix of][for detailed solving processes]{Patra2019(b)}.  Furthermore, we also calculate their $1\sigma$ errors by bootstrapping. At each bootstrap, we select a random value for each input parameter at a radius from a Gaussian distribution (the mean value is the fiducial input parameter, and the standard deviation is its uncertainty). We repeat this process 200 times and adopt the standard deviation as $1\sigma$ errors. We apply this process to all radii.

\section{Result}

\begin{figure*}
\begin{center}
	\includegraphics[width=0.8\textwidth]{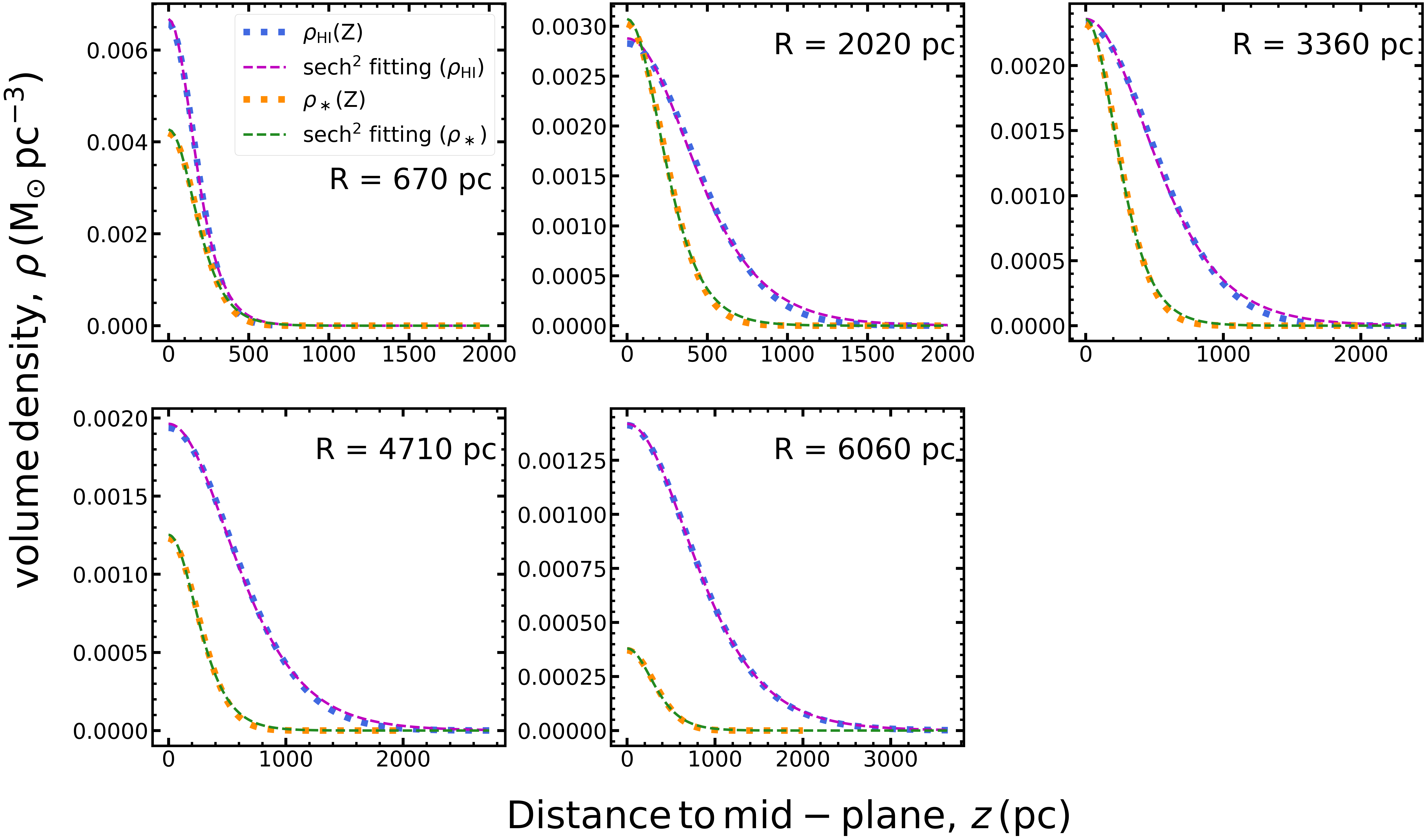}
    \caption{The derived \ion{H}{i} and stellar vertical volume density profiles of AGC 242019 at individual radii. The blue dot lines represent derived $\rho_{\rm g} (z)$, and the magenta short lines represent the corresponding fit with a sech$^2(\frac{z}{z_{\rm 0}})$ profile. The yellow dot lines represent derived $\rho_{\rm \ast} (z)$, and the green short lines for the fit with a sech$^2(\frac{z}{z_{\rm 0}})$ profile.}
    \label{fig:solution}
\end{center}
\end{figure*}

By solving the equation~(\ref{eq:final}), we obtain the vertical distributions of both stellar and gaseous volume densities. The convergent analytical solution is up to 6.06 kpc possibly because of faint stellar volume density beyond this radius. The results at each radius for AGC 242019 is shown in Fig.~\ref{fig:solution}. We fit the stellar and gas vertical profile with sech$^2(z)$:
\begin{equation}
    \rho_{\rm i}(z) = \rho_{\rm i,0} \times \frac{4}{\exp{(\frac{2z}{z_{\rm 0}})}+\exp{(\frac{-2z}{z_{\rm 0}})}+2},
\label{eq:sech2}
\end{equation}
where $\rho_{\rm i,0}$ is volume density at mid-plane, $z_{\rm 0}$ is the dispersion. We define the scale height $h_{\rm i}$ as the half-width at half maxima (HWHM) of the vertical distribution, which gives $h_{\rm i} = z_{\rm 0} \log(1+\sqrt{2})$. We further define the ratio $h_{\rm i}/R$ as the flaring angle to describe the flaring of the disk at a given radius. Since the stellar scale heights $h_{\rm \ast}$ depend on the adopted $\left \langle\sigma_{\rm z}\right \rangle_{\rm \ast}$, we focus only on the \ion{H}{i} scale height $h_{\rm g}$ in this study. In addition, while sech$^2$ is suitable to describe the gas vertical distribution, stellar disks usually deviate from sech$^2$ in observations \citep[e.g.][]{Sarkar2019}. However, we find that the changes in $h_{\rm g}$ are small when we adopt sech$^{2/n}$ distribution for the stellar distribution ($n$ is a free parameter to fit, and hardly deviates from 1 in our test).

\begin{figure*}
\begin{center}
	\includegraphics[width=\textwidth]{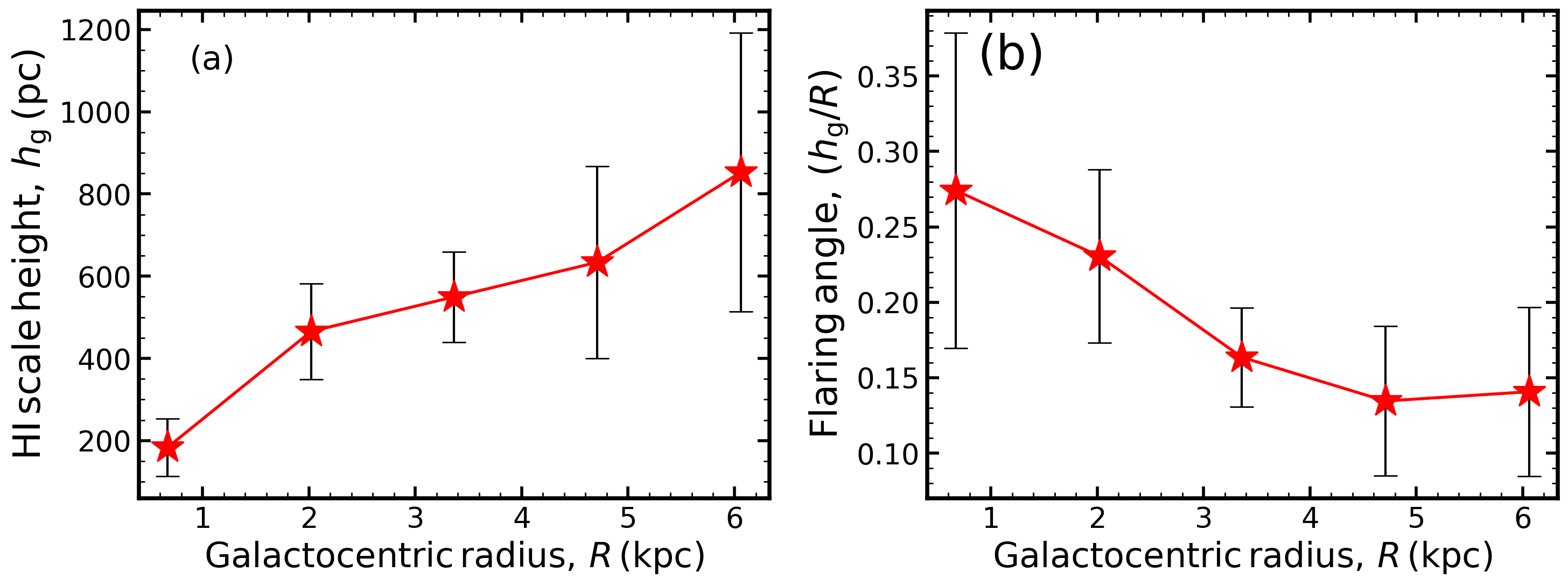}
    \caption{ The \ion{H}{i} scale heights (a) and flaring angle (b) of AGC 242019.}
    \label{fig:UDG}
\end{center}
\end{figure*}

For AGC 242019, the radial profiles of the \ion{H}{i} scale height $h_{\rm g}$ and flaring angle $h_{\rm g}/R$ are shown in Fig.~\ref{fig:UDG}, with $h_{\rm g}$ listed in Table~\ref{tab:para1}. The error bars are mostly contributed by errors of $\left \langle\sigma_{\rm z}\right \rangle_{\rm g}$. The mean \ion{H}{i} scale height ($\left \langle h_{\rm g} \right \rangle$) of AGC 242019 in the range of 0.67 to 6.06 kpc ($0.1 R_{\rm d} \sim 1.2 R_{\rm d}$) is $537.15 \pm 89.4$ pc, with standard deviation of 218.71 pc, and the mean flaring angle ($\left \langle h_{\rm g}/R \right \rangle$) is $0.19 \pm 0.03$, with standard deviation of 0.05. The intrinsic axis ratio can be defined as $2h_{\rm g}/R$, which gives the mean axis ratio of $0.38 \pm 0.06$ for AGC 242019, consistent with those of stellar disks predicted by \cite{Liao2019} and \cite{Rong2020} for isolated UDGs.

\begin{figure}
	\includegraphics[width=\columnwidth]{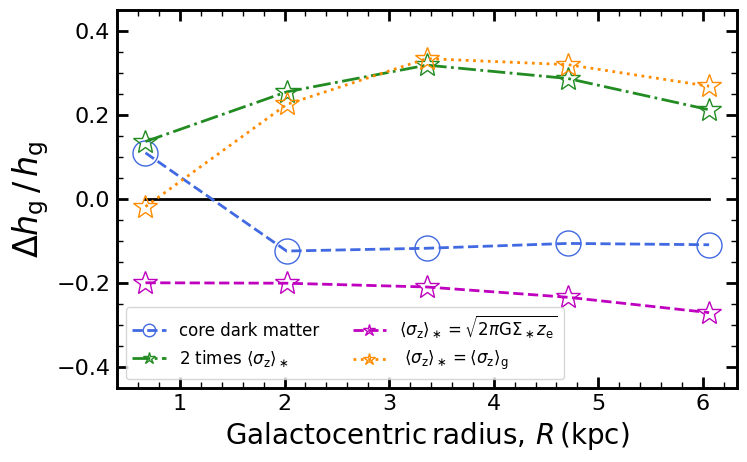}
    \caption{Differences in the gas disk scale height caused by varying the adopted stellar vertical velocity dispersion $\left \langle\sigma_{\rm z}\right \rangle_{\rm \ast}$ and dark matter profile. $\Delta h_{\rm g}$ is the difference between the $h_{\rm g}$ (with fiducial input parameters) and $h'_{\rm g}$ (with different input parameters). The blue circle and dot line represents the case with a ISO dark matter profile. The forest green star and  dot-dash line represent the case by doubling the $\left \langle\sigma_{\rm z}\right \rangle_{\rm \ast}$. The magenta stars and dash line represents the case by adopting the $\left \langle\sigma_{\rm z}\right \rangle_{\rm \ast} = \sqrt{ 2\pi {\rm G} \Sigma_{\ast} z_{\rm e}}$, where $z_{\rm e} = z_{\rm 0,\ast}/2$ is the half of the sech$^2(\frac{z}{z_{\rm 0}})$ dispersion after solving equation~\ref{eq:final}. The orange stars and dot line represents the case by adopting $\left \langle\sigma_{\rm z}\right \rangle_{\rm \ast} = \left \langle\sigma_{\rm z}\right \rangle_{\rm g}$.}
    \label{fig:test}
\end{figure}

To test the influence of the adopted stellar velocity dispersion $\left \langle\sigma_{\rm z}\right \rangle_{\rm \ast}$ on the derived \ion{H}{i} disk thickness, we use $\Delta h_{\rm g}/h_{\rm g} = (h'_{\rm g}-h_{\rm g})/h_{\rm g}$ to quantify the fractional differences  between the fiducial case ($h_{\rm g}$) in \S~\ref{sec:2}  and the case ($h'_{\rm g}$) of different $\left \langle\sigma_{\rm z}\right \rangle_{\rm \ast}$. $h'_{\rm g}$ has calculated
by the following cases: (i) $\left \langle\sigma_{\rm z}\right \rangle_{\rm \ast}$ varies by a factor of two; (2) $\left \langle\sigma_{\rm z}\right \rangle_{\rm \ast}$ is replaced by $\sqrt{2 \pi G \Sigma_{\rm \ast}  z_{\rm e}}$, $z_{\rm e} = z_{\rm 0,\ast}/2$, where $z_{\rm 0,\ast}$ is what we derive previously; (3) $\left \langle\sigma_{\rm z}\right \rangle_{\rm \ast}$ is set to be equal to $\left \langle\sigma_{\rm z}\right \rangle_{\rm g}$.  Fig.~\ref{fig:test} demonstrates that the gas disk thickness is insensitive to the adopted $\left \langle\sigma_{\rm z}\right \rangle_{\rm \ast}$, with the variation less than $33\%$. To test the influence of the adopted dark matter profile on the \ion{H}{i} scale height, we instead adopt the ISO model in equation~(\ref{eq:final}) for AGC 242019, where the ISO model parameters are provided by \cite{Shi2021}. As shown in Fig.~\ref{fig:test}, there are only less than $12\%$ differences. We also did the above two sets of tests for dwarf irregulars, which gives similar conclusions.

\begin{figure*}
\begin{center}
	\includegraphics[width=0.8\textwidth]{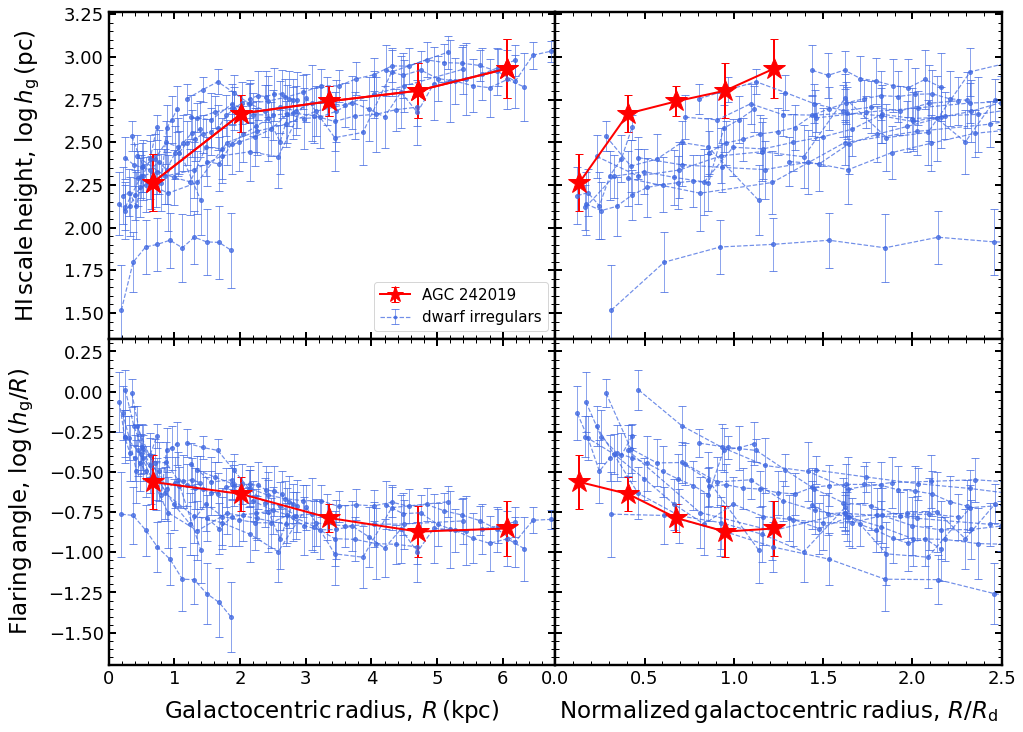}
    \caption{The \ion{H}{i} scale height (up) and faring angle (down) of AGC 242019
    as compared to those of dwarf irregulars. The left panels are the scale height ($h_{\rm g}$) and flaring angle ($h_{\rm g}/R$) as a function of the physical radius, while the right panels are as a function of the normalized radius $R/R_{\rm d}$. The red stars and solid lines represent AGC 242019, and the small blue dots and dash lines represent dwarf irregulars.}
    \label{fig:compare}
\end{center}
\end{figure*}

The radial variation of $h_{\rm g}$ of each dwarf irregular is shown in Fig.~\ref{fig:dwarf} of Appendix~\ref{sec:A1}. We give the mean $\left \langle h_{\rm g} \right \rangle$ of each dwarf irregular in the range of 0.67 to 6.06 kpc (Table~\ref{tab:dwarf_para1} (7)) and $0.1 R_{\rm d}$ to $1.2 R_{\rm d}$ (Table~\ref{tab:dwarf_para1} (8)) separately. In the range of 0.67 to 6.06 kpc, the mean $\left \langle h_{\rm g} \right \rangle$ of all dwarf irregulars is around $474.76 \pm 14.38$ pc, with standard deviation of 216.75 pc, the mean $\left \langle h_{\rm g}/R \right \rangle$ is around $0.2 \pm 0.01$, with standard deviation of 0.09, and the mean intrinsic axis ratio is around $0.40 \pm 0.01$. Our results of dwarf irregulars are consistent with previous studies \citep{Banerjee2011,Patra2020} that adopted Gaussian vertical distributions. The derived mean axis ratio is slightly less than that by \cite{Roychowdhury2010}.The differences in $h_{\rm g}$ radial profiles of some dwarfs from previous studies are caused by different kinematic data. As shown in the left panels of Fig.~\ref{fig:compare}, the \ion{H}{i} scale heights and flaring angles of AGC 242019 are within the full range of dwarfs, with comparable mean values. At the given normalized radius ($0.1 R_{\rm d}$ to $1.2 R_{\rm d}$), the mean $\left \langle h_{\rm g} \right \rangle$ of all dwarf irregulars (except DDO 54, NGC 2366, and DDO 133, whose bar radius are larger than $1.2 R_{\rm d}$) is around $262.34 \pm 11.76$ pc, with a standard deviation of 138.5 pc, the mean $\left \langle h_{\rm g}/R \right \rangle$ is around $0.37 \pm 0.02$, with a standard deviation of 0.2, and the mean intrinsic axis ratio is around $0.74 \pm 0.04$. As shown in the right panels of Fig.~\ref{fig:compare}, the mean $\left \langle h_{\rm g} \right \rangle$ ($\left \langle h_{\rm g}/R \right \rangle$) of AGC 242019 is about 0.3 dexes larger (lower) than dwarf irregulars, but still in the upper (lower) bound of dwarf irregulars within $2 \sigma$ ($1 \sigma$) scatter. Among 14 dwarf irregulars, DDO 101 has both the lowest scale heights and flaring angles shown as an outlier of Fig.~\ref{fig:compare}. This may be because dark matter, gas and stars in DDO 101 are centrally concentrated as compared to other dwarfs, which gives a steep rotation curve as well as steep radial profiles of dark matter, gas and stellar mass surface density in its inner region.

\section{Discussion}

\begin{figure*}
\begin{center}
	\includegraphics[width=\textwidth]{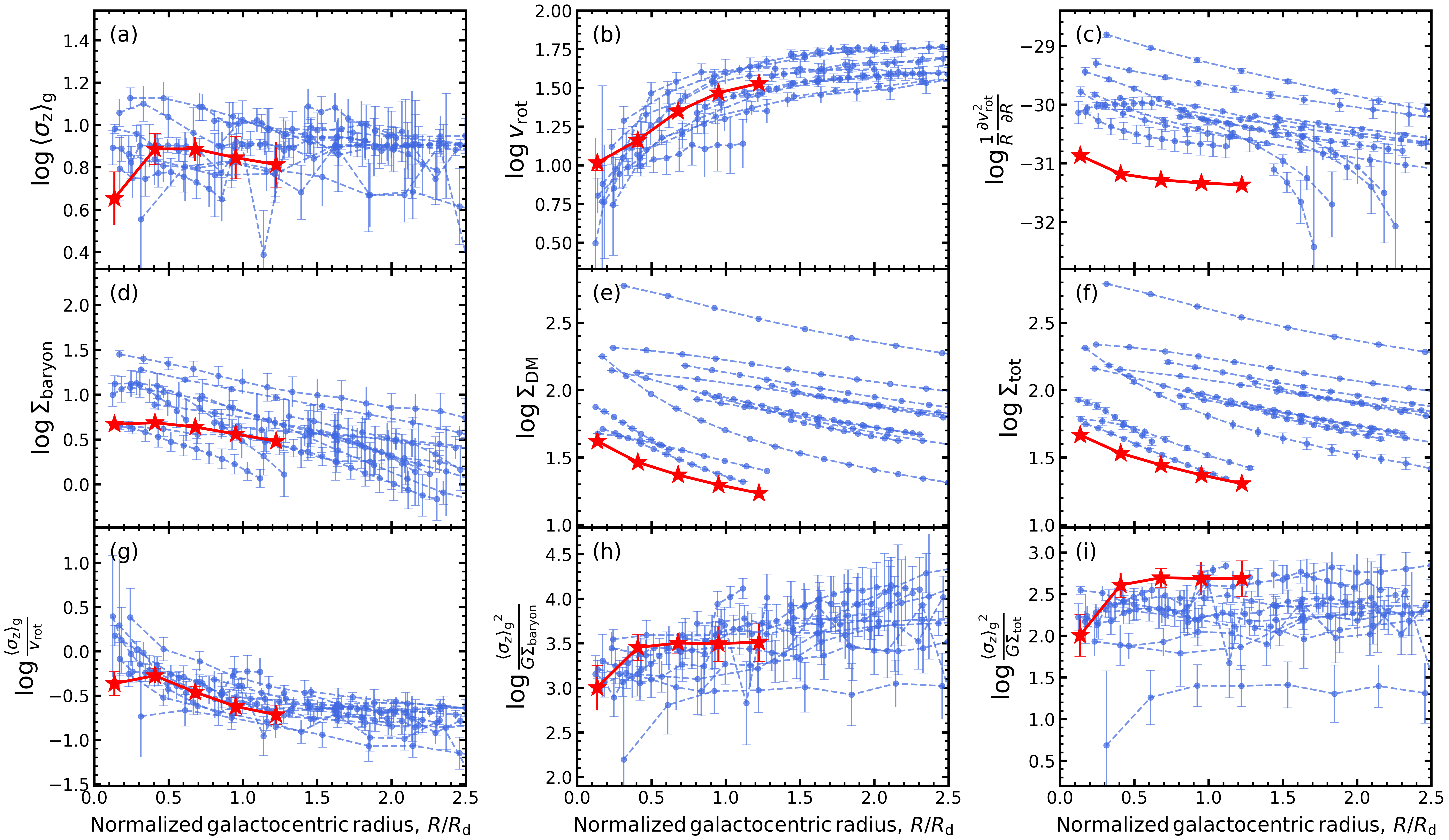}
    \caption{The physical parameters as a function of normalized radius for AGC 242019 as compared to dwarf irregulars: (a) the gas vertical velocity dispersion ($\left \langle\sigma_{\rm z}\right \rangle_{\rm g}$, in km/s); (b) the circular velocity ($v_{\rm rot}$, in km/s); (c) the radial term of gravitational potential energy ($\frac{1}{R} \frac{\partial}{\partial R}(R \frac{\partial \Phi _{\rm tot}}{\partial R})$, in ${\rm s^{-2}}$); (d) the mass surface density of baryons ($\Sigma_{\rm baryon}=\Sigma_{\rm \ast}+\Sigma_{\rm gas}$, in ${\rm M_{\odot}/{pc}^2}$); (e) the mass surface density of dark matter ($\Sigma_{\rm DM} = 2\int_{0}^{\infty}\rho_{\rm DM}(R,z)dz$, in ${\rm M_{\odot}/{pc}^2}$) ; (f) the total mass surface density ($\Sigma_{\rm tot}=\Sigma_{\rm baryon}+\Sigma_{\rm DM}$, in ${\rm M_{\odot}/{pc}^2}$); (g) the ratio of the vertical velocity dispersion and rotation velocity ($\left \langle\sigma_{\rm z}\right \rangle_{\rm g}/v_{\rm rot}$); (h) ${\left \langle\sigma_{\rm z}\right \rangle_{\rm g}}^2/G\Sigma_{\rm baryon}$, in pc; (i) ${\left \langle\sigma_{\rm z}\right \rangle_{\rm g}}^2/{\rm G}\Sigma_{\rm tot}$, in pc. The red stars and solid lines are AGC 242019, while the blue dots and dash lines are dwarf irregulars.}
    \label{fig:properties}
\end{center}
\end{figure*}

To further understand what drives similar scale heights between AGC 242019 and dwarf irregulars, we plot individually logarithmic parameters in equation~(\ref{eq:final}) as a function of the  radius normalized by $R_{\rm d}$ in Fig.~\ref{fig:properties}: (a) the gas vertical velocity dispersion ($\left \langle\sigma_{\rm z}\right \rangle_{\rm g}$); (b) the circular velocity ($v_{\rm rot}$); (c) the radial term of gravitational potential energy ($\frac{1}{R} \frac{\partial}{\partial R}(R \frac{\partial \Phi _{\rm tot}}{\partial R})$); (d) the mass surface density of baryon ($\Sigma_{\rm baryon}=\Sigma_{\rm \ast}+\Sigma_{\rm gas}$); (e) the mass surface density of dark matter ($\Sigma_{\rm DM} = 2\int_{0}^{\infty}\rho_{\rm DM}(R,z)dz$) ; (f) the total mass surface density ($\Sigma_{\rm tot}=\Sigma_{\rm baryon}+\Sigma_{\rm DM}$); (g) the ratio of the vertical velocity dispersion and the rotation velocity ($\left \langle\sigma_{\rm z}\right \rangle_{\rm g}/v_{\rm rot}$) that determines the turbulent motions relative to regular motions; (h) ${\left \langle\sigma_{\rm z}\right \rangle_{\rm g}}^2/{\rm G}\Sigma_{\rm baryon}$ that describes the balance between pressure and the gravity of baryons; (i) ${\left \langle\sigma_{\rm z}\right \rangle_{\rm g}}^2/{\rm G}\Sigma_{\rm tot}$ that describes the balance between pressure and the total gravity. 

AGC 242019 has comparable gas vertical velocity dispersion, circular velocity and the ratio of the two as dwarf irregulars, as shown in Fig.~\ref{fig:properties} (a), (b) and (g), respectively. The baryonic mass surface density, the dark matter mass surface density and the
sum of the two of AGC 242019 is lower than the majority of dwarf irregulars as shown in Fig.~\ref{fig:properties} (d), (e) and (f), respectively. The radial gravitational potential is obviously lower than dwarf irregulars too as shown in Fig.~\ref{fig:properties} (c). The low radial gravitational potential reflects a slowly rising rotation curve, as a result of low mass surface densities. From Fig.~\ref{fig:properties} (h) and (i), the balance between turbulence and gravity of AGC 242019 is still within the range of dwarf irregulars. In fact, $h_{\rm g} \propto {\left \langle\sigma_{\rm z}\right \rangle_{\rm g}}^2/{\rm G}\Sigma$ is a simplified form of hydrostatic equilibrium whose coefficients depend on the adopted disk models \citep[e.g.][]{KS1,KS2,Kruit1981(99),Kruit1988, Bottema1993, Kregel2002, Ben2018, Wilson2019, Das2020}. From Fig.~\ref{fig:properties} (h) to Fig.~\ref{fig:properties} (i), AGC 242019 relative to dwarf irregulars have slight increase in ${\left \langle\sigma_{\rm z}\right \rangle_{\rm g}}^2/{\rm G}\Sigma$ after including dark matter in $\Sigma$, implying the insignificance of dark matter in AGC 242019 relative to that in dwarfs.

\begin{figure}
	\includegraphics[width=\columnwidth]{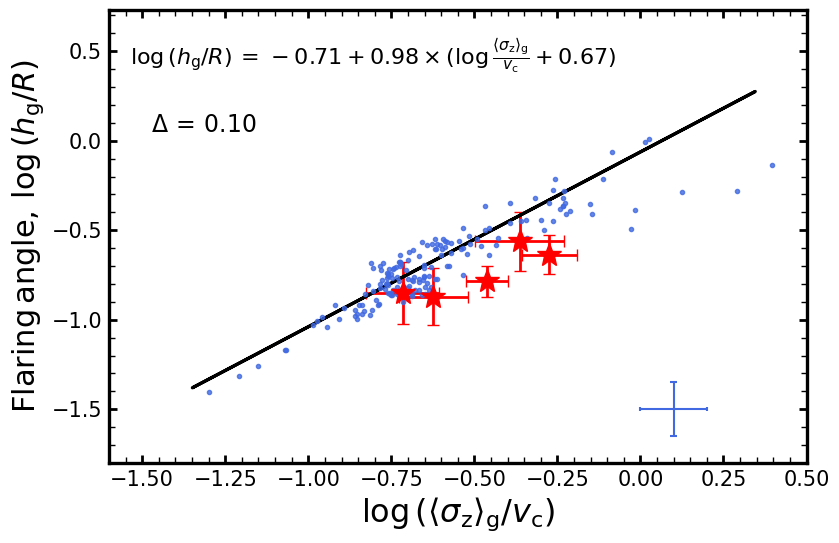}
    \caption{$h_{\rm g}/R$ versus ${\left \langle\sigma_{\rm z}\right \rangle_{\rm g}}/{v_{\rm c}}$. The red stars are  AGC 242019 and blue dots are dwarf irregulars. The median errors for dwarf irregulars are labelled in the bottom right corner of the figure. The best linear fitting is ${\rm log}\, (h_{\rm g}/R) = (0.977 \pm 0.064)({\rm log}\, (\left \langle\sigma_{\rm z}\right \rangle_{\rm g}/{v_{\rm c}})+0.666)- (0.713 \pm 0.013)$ with standard deviation of $0.10$ dex.}
    \label{fig:vd_to_vc}
\end{figure}

By combining all objects together, we find that $h_{\rm g}/R$ is related with $\left \langle\sigma_{\rm z}\right \rangle_{\rm g}/v_{\rm rot}$ as shown in Fig.~\ref{fig:vd_to_vc} \footnote{Fitting with package LtsFit \citep{Cappellari2013}.}. This can be roughly derived from $h_{\rm g} \propto {\left \langle\sigma_{\rm z}\right \rangle_{\rm g}}^2/{\rm G}M_{\rm dyn}$ and ${\rm G}M_{\rm dyn}={v_{\rm c}}^2/R$. At $\left \langle\sigma_{\rm z}\right \rangle_{\rm g}/v_{\rm rot} > 1$ there is a large scatter, which may be caused by large uncertainties of kinematic measurements in the central regions of DDO 53 and DDO 50. In spite of these outliers, the correlation is tight with a similarity to the observed relation for stellar disks of spirals in the form of ${(h_{\rm g}/R)}_{\rm R=R_{\rm d}} \propto \left \langle\sigma_{\rm z}\right \rangle_{\rm \ast}/{v_{\rm c,max}}$  \citep[e.g.][]{Bottema1993,Kregel2002,Kaufmann2007, Janssen2010}. This implies a similar origin of the flaring of gaseous disks to that of stellar disks, which is the hydrostatic equilibrium under
self gravity. 

AGC 242019 and dwarf irregulars are isolated systems so that their disk thickness is not likely  caused by environmental effects, although some dwarfs show faint tidal tails \citep[i.e][]{Hunter2006,Zhang2012}. On the other hand, the stellar feedback is found to be important for dwarf galaxies: \cite{Hunter1993} found that NGC 2366, DDO 50, and DDO 53 have shell structures; \cite{Egorov2021} identified six expanding superbubbles in DDO 53; DDO 50 contains many \ion{H}{i} holes \citep{Puche1992, Iorio2017}; \citet{Zhang2012} derived star formation histories (SFH) with the SED fitting method and found that most of them have went through active star formation phases. Stellar feedback is also proposed to be important to expand the disk sizes
for \ion{H}{i}-bearing UDGs \citep[][]{Jiang2019,Liao2019}, which could thicken the disk as well
\citep[][]{Janssen2010,Di_Cintio2017,Jiang2019,Liao2019}. 
This study indicates that AGC 242019 has a \ion{H}{i} scale height comparable to dwarf irregulars.
This implies that AGC 242019 should not experience much stronger feedback than dwarf galaxies.  This is consistent with other evidences as found in \cite{Shi2021}. 

\section{Conclusion}
In this study, we derive the radial profiles of \ion{H}{i} scale height and flaring angle of AGC 242019 and a sample of 14 dwarf irregulars for comparison. We model a galaxy as a two-component isothermal and axisymmetric disk in a dark matter halo. Assuming a self-gravitating system in hydrostatic equilibrium, we solve the joint Poisson-Boltzmann equation to estimate the gas density vertical distribution, and define the HWHM of the sech$^2(\frac{z}{z_{\rm 0}})$  vertical profile as the \ion{H}{i} scale height:

(1) In the range of $0.67$ to $6.06$ kpc ($0.1 R_{\rm d}$ to $1.2 R_{\rm d}$), AGC 242019 has a mean \ion{H}{i} disk scale height of $537.15 \pm 89.4$ pc and a mean flaring angle of $0.19 \pm 0.03$. They are comparable to those of 14 dwarf irregulars in the same physical radial range. However, in the range of $0.1 R_{\rm d}$ to $1.2 R_{\rm d}$, the mean \ion{H}{i} disk scale height of AGC 242019 is 0.3 dex larger, and the mean flaring angle is 0.3 dexes smaller than the dwarf irregulars, but still in the upper ($2 \sigma$) and lower ($1 \sigma$) bound of dwarf irregulars, respectively.

(2) The \ion{H}{i} disk height of AGC 242019 is overall within the range of dwarf irregulars. This implies that AGC 242019  unlikely has experienced much stronger feedback than dwarf irregulars. 

(3) The \ion{H}{i} disks of AGC 242019 and dwarf irregulars follow a relation of  ${\rm log}\, (h_{\rm g}/R) = (0.977 \pm 0.064)({\rm log}\, (\left \langle\sigma_{\rm z}\right \rangle_{\rm g}/{v_{\rm c}})+0.666)- (0.713 \pm 0.013)$ with standard deviation of $0.10$ dex.

\section*{Acknowledgements}

We  thank  the referee  for  helpful  and constructive  comments  that improved the  paper significantly.
XL and YS acknowledges the support from the National Key R \& D Program of China (No. 2018YFA0404502, No. 2017YFA0402704), the National Natural Science Foundation of China (NSFC grants 11825302, 12141301, 12121003, 11733002), and the China Manned Space Project with NO. CMS-CSST-2021-B02. YS thanks the Tencent Foundation through the XPLORER PRIZE.

\section*{Data Availability}
The data underlying this article are available in the article.

\bibliographystyle{mnras}
\bibliography{mreference} 

\appendix

\section{}
\label{sec:A1}
\begin{figure*}
\begin{center}
	\includegraphics[width=\textwidth]{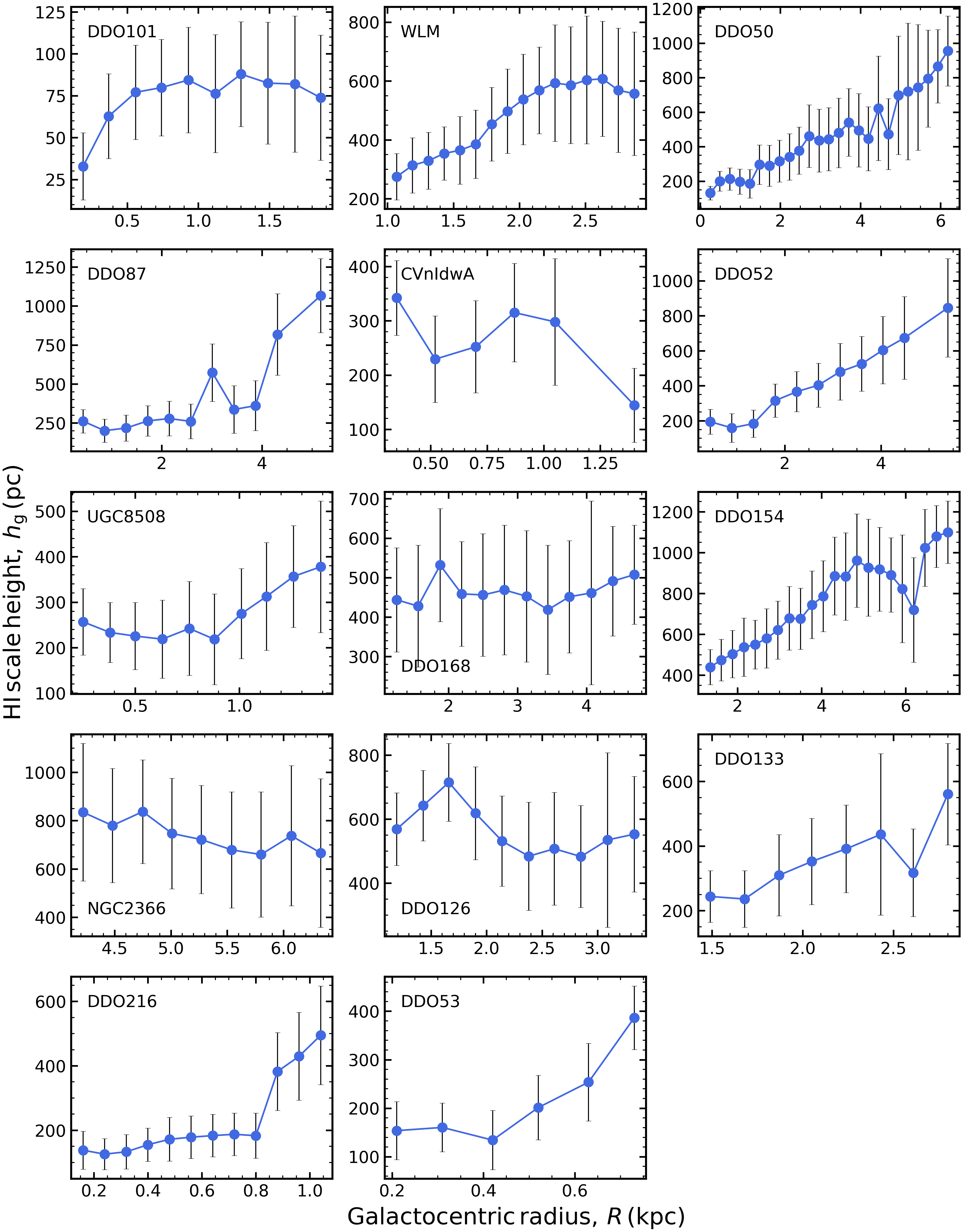}
    \caption{ The \ion{H}{i} scale heights of dwarf irregulars.}
    \label{fig:dwarf}
\end{center}
\end{figure*}

\bsp	% typesetting comment
\label{lastpage}
\end{document}